# Improvement of the superconducting properties of polycrystalline FeSe by silver addition

E. Nazarova[1,2], N. Balchev[1], K. Nenkov[2,4], K. Buchkov[1,4], D. Kovacheva[3], A. Zahariev[1] and G. Fuchs[2]

[1]*Institute of Solid State Physics, Bulgarian Academy of Sciences, 1784 Sofia, Bulgaria*
[2]*Leibniz Institute for Solid State and Materials Research, (IFW Dresden), P.O. Box 2700116, D-01171 Dresden, Germany*
[3]*Institute of General and Inorganic Chemistry, Bulgarian Academy of Sciences, 1113 Sofia, Bulgaria*
[4]*International Laboratory of High Magnetic Fields and Low Temperatures, Gajowicka 95, 53-529 Wroclaw, Poland*

E-mail:

**Abstract** We investigated the influence of different Ag additions (up to 10 wt %) on the superconducting properties of $FeSe_{0.94}$. The structural investigations (XRD and SEM) indicated that Ag is present in three different forms. Ag at grain boundaries supports the excellent intergrain connections and reduces $\Delta T$ to values smaller than 1K at B=0 and $\Delta T \leq 2.74$ K at B=14 T. Ag insertion in the crystal lattice unit cell provides additional carriers and changes the electron hole imbalance in $FeSe_{0.94}$. This results in an increase in the magnetoresistive effect (MR) and critical temperature ($T_c$). Reacted Ag forms a small amount (~1%) of $Ag_2Se$ impurity phase, which may increase the pinning energy in comparison with that of the undoped sample. The enhanced upper critical field ($B_{c2}$) is also a result of the increased impurity scattering. Thus, unlike cuprates Ag addition enhances the $T_c$, $B_{c2}$, pinning energy and MR making the properties of polycrystalline $FeSe_{0.94}$ similar to those of single crystals.

## 1. Introduction

The Ag addition is a technique used often to improve the properties of superconducting materials. In granular cuprates, Ag usually does not enter the crystallographic structure, nor forms impurity phases, but is randomly distributed at grain boundary regions. This transforms a part of the superconductor – insulator – superconductor intergrain connections in superconductor – normal metal – superconductor, which are more stable in a magnetic field. As a result, the intergranular critical current in a magnetic field is enhanced. In the low temperature A-15 phase superconductors, additional element may be present as unreacted in elemental form or in the form of separate phase (after reaction) improving the pinning [1]. Obtaining the low temperature superconductor $Nb_3Sn$ through the diffusion reaction between Nb and Ag-Sn alloys shows that Ag is not as effective in accelerating of superconductor rate formation as is Cu (from Cu-Sn) [2]. However, Ag-addition in the other intermetalic superconducting compound $MgB_2$ was found to promote the diffusion reaction of Mg, lower the reaction temperature, improve crystallinity and grain connections, which results in an enhancement of $J_c$ [3]. The Ag addition in the recently discovered iron based superconductors is yet to be studied in detail. The role of Ag addition in the structure and SC properties was investigated for polycrystalline $Sr_{0.6}K_{0.4}Fe_2As_2$ [4, 5] and $FeSe_{0.5}Te_{0.5}$ [6]. It was established that silver improves the intergrain connections and enhances the critical current $J_c$, irreversibility field $H_{irr}$ and the hysteresis magnetization of these superconductors. Moreover, a small amount of Ag was found to enter into the crystal structure of $FeSe_{0.5}Te_{0.5}$. We established that the Ag addition in $FeSe_{0.94}$ improves both the intra- and inter-granular superconducting properties and slightly increases the irreversibility field [7]. Further investigations showed [8] that critical temperature, the magnetoresistance, the upper critical field and the activation energy for thermally activated flux flow also increased as a result of 4 wt % Ag doping. This study aims to determine the mechanism of action of silver on the superconducting properties and to optimize its content in the superconducting $FeSe_{0.94}$.

## 2. Experiments

A series of samples is synthesized with nominal composition $FeSe_{0.94}$ and different Ag addition (0, 4 wt %, 6 wt %, 8 wt % and 10 wt %). For brevity, they are marked in the text below as FeSeAg0, FeSeAg4, FeSeAg6, FeSeAg8 and FeSeAg10 respectively. The preparation procedure has two steps: a solid state reaction of the initial products at $700^0$ C and subsequent melting of the obtained material at 1050-1060 C. All procedures (weighing, grinding, homogenization, pressing) are carried out in a glove box with Ar atmosphere; the syntheses are conducted in evacuated quartz ampoules (~$10^{-4}$ tor). More details for the sample's preparation are given elsewhere [8].

Powder X-ray diffraction patterns are collected within the range from 5.3 to 80° 2θ at a constant step 0.02° 2θ on a Bruker D8 Advance diffractometer with Cu Kα radiation and LynxEye detector. Phase identification is performed by the Diffrac*plus* EVA using ICDD-PDF2 Database. Microstructural and compositional investigations are performed by a dual-beam scanning electron focused ion beam system (SEM/FIB LYRA I XMU, TESCAN), equipped with an EDX detector (Quantax 200, Bruker).

The resistivity of the samples at different magnetic fields is measured by the standard four probe method using DC resistivity option of PPMS-14 T. The samples used for these measurements have a parallelepiped shape.

3. **Results and discussion**

The XRD analyses show that besides the presence of large amount of tetragonal phase in melt growth FeSeAg0 a small amount non superconducting hexagonal phase is also detected (Fig.1). This is in agreement with a previous study of FeSe phase diagram [9] where it is shown that hexagonal $\delta(Fe_{1+x}Se)$ phase (with Ni-As structure) crystallize from the melt and subsequently is transformed to the tetragonal $\beta(Fe_{1+x}Se)$ (with PbO structure). However, some residual high temperature phase is always present in the prepared samples. The Ag addition reduces the amount of the hexagonal phase below the sensibility of XRD analysis. A small increase in the intensity of (00l) peaks of $\beta(Fe_{1+x}Se)$ phase is observed, which is an indication of enhanced texturing. Except for the dominating tetragonal phase, a small quantity of $Ag_2Se$ (about 1%) and traces of Ag (Fig.

1) are established also. Non-reacted Ag is found to be present in sample FeSeAg4, while for higher Ag concentration, the impurity $Ag_2Se$ phase is seen.

Using SEM with EDX analysis we found grains of several μm consisting of more than 90 wt % Ag in sample FeSeAg4. In the sample with the largest Ag concentration (FeSeAg10) the number of grains consisting of Ag increases. They have different size ranging from few μm to more than 10 μm.

When comparing the 00l peaks of tetragonal phase for FeSeAg0 and FeSeAg10 (or FeSeAg6) samples a noticeable systematic shift of all peaks is observed (shown in Table 1). An XRD peaks displacement smaller by one order of magnitude is reported by Sudesh et all [10] and Pandya et all [11] in $FeSe_{1-x}Sb_x$. This is reasonable because the atomic radius of Ag is larger than that of Sb (160 and 145 pm respectively). The observed peaks displacement in [10] is attributed to the Sb substitution for Se. In our case, it could be an indication for incorporation of Ag in the unit cell. It turns out that in the investigated samples the Ag is present as not reacted, in $Ag_2Se$ phase and incorporated in the crystal structure. The integral EDX analysis of a homogenous region of the FeSeAg8 sample shows weakly Se-deficient iron selenide and few percent of Ag. The elements map of this region indicates an almost homogenous distribution of Fe, Se and Ag.

Fig. 2 presents the resistivity vs temperature dependences ($\rho_0(T)$) for all samples at B=0. It is seen that the normal state resistivity gradually decreases in the samples as Ag is added due to parallel paths formation at the grain boundaries [8]. The residual resistance ratio (RRR=R(T=300K)/R(T=16 K)) increases from 2.95 for undoped sample to 9.74 for the sample FeSeAg6 indicating an improvement of the sample's purity (see Table 2). This could be a result of a suppression of the hexagonal phase formation. For the sample with the highest Ag concentration (FeSeAg10), RRR decreases to 3.25 probably due to the relatively enhanced amount of the $Ag_2Se$ impurity phase. The inset shows the superconducting transitions where for the purpose of a comparison the values of $\rho_0$ for sample FeSeAg0 are reduced by factor of 2. The critical temperature $T_c(\rho_0=0)$ increases from 7.3 K for undoped sample to 9.05 K for the sample FeSeAg6. The $T_c$ enhancement with ~2K could be an indirect evidence of Ag incorporation in the crystal structure. The detected shifts of the XRD (00l) peaks to the smaller angles in Ag doped sample presuppose a little expansion of the **c** distance in the unit cell. This situation is opposite to the collapse of $Fe_2Se_2$ layers separation under pressure when $T_c$ in $\beta$-$Fe_{1.01}Se$ increases from 8.5 to 36.7 K under pressure of 8.9 GPa [12]. It is reasonable to expect that in our samples the influence of the superconducting transition temperature is a result of doping and this will be discussed when the magnetoresisive effect is described. The width of superconducting transition ($\Delta T$) is determined according to 10-90 % criterion. $\Delta T(B=0)$ is found do decrease from 2.27 K for the undoped sample to 0.70 K for sample FeSeAg8 and in spite of slight increase (to 0.94 K for sample FeSeAg10) remains below 1 K for all doped samples. In a magnetic field $\Delta T$ increases. The weak links behavior in our samples is less pronounced, especially for the Ag doped ones. In a magnetic field of 14 T, $\Delta T \leq 2.74$ K for the doped samples and $\Delta T = 3.63$ K for the undoped one. For sample FeSeAg6, the transition width increases by about 1 K when the magnetic field is raised from 0 to 14 T i.e. $\Delta T(B=14T) - \Delta T(B=0) = 1.07$ K. Probably in this sample the excellent intergrain connections are due to the current path entirely through the Ag coupled grains and the Ag thickness is of the order of coherence length. For smaller and higher Ag concentrations these optimal conditions are not achieved.

Comparing our results with those existing in the literature for Si, Sb [10,11] and Sn [13] doping in FeSe we should mention that the Ag addition reduces significantly the normal state resistivity in samples with Ag addition, while the opposite effect is observed from the other authors. No matter whether the resistance in normal state increases (or decreases) in both cases the samples become superconducting. This implies the absence of the conventional coupling mechanism whereby the high normal state resistance is compatible with conventional superconductivity.

A structural phase transition from tetragonal (at room temperature) to orthorhombic phase (below 80 K) exists in FeSe [14]. This structural transition occurs in all investigated samples, which is well seen in the temperature dependences of the resistivity derivatives presented in the inset of Fig.1. The temperature of this structure transformation does not change by Ag doping with different concentrations.

Worth mentioning is the linear temperature dependence of the resistivity in normal state for all samples in the temperature interval determined by the superconducting (~10K) and structural phase transitions (~80K). This behavior is similar to that of cuprates above $T_c$. This is a characteristic sign for non Fermi liquid behavior and strong electron correlations in both types of high temperature superconductors (Fe-based and cuprates). Non-Fermi-liquid behavior is often found near a magnetically ordered phase in the phase diagram, indicating that the non-Fermi-liquid state in those systems may be linked to a magnetic instability arising at $T=0$ [15].

The resistivity vs. temperature dependences ($\rho_H(T)$) are investigated for all the samples at different magnetic fields up to 14 T. A moderate magnetoresistance effect (MR) is observed in samples with Ag addition (see Table 1). In the previous work [8] only one sample with 4% Ag addition was examined, so that we assumed that the registered MR (~50%) is due to Ag and $Ag_2Se$ impurity phases. However we established here that as the Ag addition is increased, the most intensive peak of $Ag_2Se$ (at $2\theta=33.525$) becomes a little bit stronger, while the MR is weakly suppressed (between 34% and 39% – Table 1). Furthermore, the sample without Ag (FeSeAg0) also shows a small MR effect. These contradictions imply that the MR effect in our samples may have another origin.

According to the theoretical calculations [16] and different experimental investigations [17-19], a multiband electronic structure is assumed to be realistic for the FeSe superconductor. This brings to mind the other multiband superconductor $MgB_2$, where a large normal state MR was observed and studied in details (see for example [20]). It has been shown that for a single band free electron system a zero MR is predicted. However, in multiband solid, where each band has different average cyclotron frequency ($\omega_c$) and relaxation time ($\tau$) large MR is expected. Another indication of the multiband effect is the failing of Kohler's rule. The latter states that if only one type of carriers exist with one relaxation time, MR should be a function of $B/\rho_0$ and MR at different temperatures should collapse to a single curve in coordinates ($\rho_H - \rho_0$)/ $\rho_0$ versus $B/\rho_0$ (Kohler plot). Coming back to the FeSe investigated here, we should point out that the stoichiometric compound has an equal number of electrons and holes [18], which presuppose an absence of MR. However, this balance is disturbed in the investigated Se deficient iron selenide ($FeSe_{0.94}$) and the observed small MR in FeSeAg0 sample could be a result of electrons and holes imbalance. The Ag insertion in the crystal lattice unit cell could change this imbalance. To support this assumption Fig.3 presents the dependence $T_c$ vs. Ag concentration. The $T_c$ is determined from the temperature dependent resistive transition (at B=0) at the point $0.5\rho_n$. The domelike shaped dependence obtained is similar to that observed in cuprates when the $T_c$ is presented as a function of the increasing carrier concentration. Thus, the Ag addition changes the carrier concentration which results in a significant MR effect in doped samples. In the investigation already cited [11], we can see notable MR also for x=0.05 and 0.10 Sb or Si substitutions.

We demonstrated that a Kohler plot is found for the sample FeSeAg4 in the temperature interval 11K – 20K [8]. However, the additional magnetoresistive measurements of this sample at higher temperatures (up to 90 K) show that the Kohler plot fails. The experimental results obtained are presented in Fig.4. For the sample FeSeAg8, a Kohler plot does not exist in the temperature interval Tc – 80 K either. These results support the multiband structure of our samples and agree with their non Fermi liquid behavior in the temperature interval mentioned. The scaling observed at low temperatures may have a more complicated explanation depending on the changes in the mobility of the different types of carriers in this case [20]. On the other hand, the existence of a Kohler plot has been found for the narrow gap semiconductor $Ag_2Se$ in the broad temperature interval 50K – 150K [21]. This means that the MR observed here does not originate from the formed $Ag_2Se$ in the samples; moreover its content is determined to be very small.

In Fig. 5, upper critical field data of all samples are compared using the midpoint of the resistive transition at $0.5\rho_n$ (with $\rho_n$ as the normal state resistivity) to define the upper critical field $B_{c2}$. Due to the Ag addition, the $B_{c2}^{50}(T)$ line of the undoped sample is found to shift to higher fields. The $B_{c2}(T)$ lines almost coincide for all samples with Ag except for the highest Ag concentration (FeSeAg10), where the $B_{c2}(T)$ line falls down to lower fields, but still remains higher than that of the undoped sample. Similar to the other multiband superconductor, $MgB_2$ [22], an upward curvature is present in the $B_{c2}(T)$ dependence close to $T_c$ which transforms to linear at lower temperatures.

In Fig. 6, $B_{c2}(T)$ data for FeSe samples, undoped and doped with 4, 6 and 8wt % Ag, are fitted by the WHH approach [23] for one band BCS superconductors in the dirty limit. The fit curves for the upper critical field $B_{c2}^*(T)$ related to the orbital pair breaking effect are shown as black lines in Fig. 6. They start with a linear slope $dB_{c2}/dT$ from $T_c^* < T_c$, i.e. in order to fit the data, the positive curvature of $B_{c2}(T)$ near $T_c$ was neglected. At $T = 0$, the orbital upper critical field $B_{c2}^*(0)$ is given by

$$B_{c2}^*(0) = -0.69\, T_c^* |dB_{c2}/dT|_{T_c^*}. \qquad (1)$$

The $B_{c2}(T)$ data of all three Ag doped samples are fitted by a common fit curve which describes the $B_{c2}(T)$ data for these samples in a restricted field range up to about 5-7 T. The $B_{c2}(T)$ data at higher fields exceed the orbital $B_{c2}^*(T)$ which is a clear indication for multiband superconductivity in the FeSe samples investigated.

At sufficiently high magnetic fields, the superconductivity is destroyed by orbital pair-breaking and spin pair-breaking. Within the WHH approach, the Maki parameter

$$\alpha = \sqrt{2}\, B_{c2}^*(0)/ B_p(0) \qquad (2)$$

provides a convenient measure for the relative strength of orbital and spin pair-breaking.

Here, $B_p(0)$ is the Pauli limited field [24]:

$$B_p(0) = 1.86 (1+\lambda) T_c^* \tag{3}$$

with $\lambda$ as electron-boson coupling constant. Using $\lambda \sim 1.5$ as representative value [25], one estimates high values of $B_p(0)$ exceeding the orbital field $B_{c2}^*(0)$ (see Table 3). Using the Maki parameters obtained of $\alpha = 0.82$ and 0.92 for the undoped and Ag doped samples, respectively, a flattening of $B_{c2}(T)$ is expected at high fields due to Pauli limiting as shown in Fig. 6. This flattening of $B_{c2}(T)$ should be more pronounced for the Ag doped FeSe samples. Unfortunately, the expected flattening of $B_{c2}(T)$ is masked by the upturn of $B_{c2}(T)$ associated with multiband superconductivity in the available field range.

However, a similar shape of $B_{c2}(T)$ has been recently reported for a FeSe single crystal for magnetic fields $H \parallel ab$ which was investigated up to high magnetic fields [26]. In Fig. 7, $B_{c2}(T)$ data of that FeSe single crystal and Ag doped FeSe samples are compared. Both have a comparable $T_c$, but the slope of $B_{c2}(T)$ of the single crystal near $T_c$ is, with $dB_{c2}/dT \sim -6.2$ T/K, larger than that of the Ag doped samples by a factor of 1.4. In order to compare the shape of $B_{c2}(T)$ data, the $B_{c2}(T)$ data for the single crystal for fields $H \parallel ab$ are scaled with $1/1.4 \sim 0.7$. The scaled data points obtained for the single crystal (for $H \parallel ab$) shown in Fig. 7 are well described by the WHH model using $\alpha = 0.92$. Only a slight upturn of the scaled $B_{c2}(T)$ data around 6 K is visible for the single crystal indicating multiband superconductivity, but at lower temperatures the flattening of $B_{c2}(T)$ due to Pauli limiting becomes dominant.

It is demonstrated that by doping polycrystalline FeSe with Ag, a similar shape of $B_{c2}(T)$ is obtained as for single crystals for magnetic fields $H \parallel ab$ which shows a flattening of $B_{c2}(T)$ due to Pauli limiting. Generally, the $B_{c2}$ values derived in the polycrystalline samples from the onset of the superconducting transition refer to those grains which are oriented with their $ab$ planes along the applied field. Using this definition of $B_{c2}$, the slope of $B_{c2}$ for the Ag doped samples would increase up to about $dB_{c2}/dT \sim -5.5$ T/K which is not far from the slope of $dB_{c2}/dT \sim -6.2$ T/K for the single crystal for fields $H \parallel ab$.

The critical current density is another important characteristic of superconducting materials. Its fundamental limit is determined from the depairing current density [27]:

$$J_d = (2/3)^{3/2} H_c / \lambda , \tag{4}$$

where $H_c$ is the thermodynamic critical magnetic field and $\lambda$ is the penetration depth. At this value the kinetic energy of the superconducting carriers exceeds the binding energy of Cooper pairs. Using previously obtained values for $\lambda$, k and $H_{c2}(0)$ ($H_{c2}(0) = H_c/\sqrt{2}.k$) [8] we estimate the depairing critical current density for sample FeSeAg4 to be of the order of $10^{11}$ A/m$^2$. The value of the same order has been obtained by using the corresponding data for a single crystal [28]. This value is close to the depairing critical current density of the other superconductors, namely $10^{12} - 10^{13}$ A/m$^2$ [29]. However, in the real superconducting materials the critical current density is always smaller. Using the temperature dependence of imaginary part of the fundamental AC magnetic susceptibility, $\chi_1''(T)$, and the sample's dimensions we estimate the intergranular critical current density for sample FeSeAg6 to be about $3.5.10^5$ A/m$^2$ at 4.5 K. On the other hand, even in the multidomain FeSe crystals the critical current does not exceed $6.10^7$ A/m$^2$ at 4 K and H=0.5 T [30]. The critical current density in our polycrystalline sample is strongly suppressed for several reasons. The first one is the low mass density of investigated sample, which is ~70% from the theoretically calculated value – 5.66 g/cm$^3$; secondly – the presence of non superconducting phases even at small quantity and finally the presence of not fully oriented crystallites. Additional technological work is needed for further improvement of these characteristics.

Fig. 8 shows the $\ln(U/k_B)$ vs $\ln B$ plots for the samples investigated. The values of $U/k_B$, where $k_B$ is the Boltzmann constant, are obtained from the slopes of the $\ln\rho$ vs $1/T$ dependences at the different measuring fields. The activation energy $U$ shows a weak field dependence at low fields, changing in a stronger one at $B \sim 4T$. Both branches can be described by a power law behavior, $U \sim B^{-\beta}$, as indicated by the straight lines in this double-logarithmic plot. The parabolic-like shape of the curves, mostly expressed in the undoped sample may be explained by the presence of strong grain boundary pinning [11]. At low fields (B < 4T) β increases from 0.15 for FeSeAg0 to 0.48 - 0.57 for the Ag doped samples. At high fields ($B > 4T$), the field dependence of $U$ becomes stronger, i.e. β increases to 1.65 - 1.9 for all the investigated samples. A similar field dependence of $U$ is reported for *β*-FeSe [28] and Fe(Te,S) single crystals [31] and is associated with a crossover from single-vortex pinning at low magnetic fields to collective pinning at high fields. According to Fig. 8, Ag doping results in a strong enhancement of the activation energy for thermally activated flux flow at low fields and reaches maximal value for FeSeAg6. This is not observed in Si and Sb doped FeSe$_{0.9}$, where $U/k_B$ decreases with the increase of the Si and Sb content. The enhancement of $U$ in our case may be due to the presence of Ag and/or Ag$_2$Se impurities in the bulk of the grains, which may act as pinning centers.

## 4. Conclusion

We investigated the superconducting properties of polycrystalline FeSe$_{0.94}$ samples doped with different (0, 4, 6, 8 and 10 wt %) Ag concentration. The structural investigations indicate that Ag suppressed the formation of the non superconducting hexagonal phase. Ag is present as unreacted, incorporated in the crystal structure and as a small amount (~ 1%) of Ag$_2$Se phase. The Ag-doping increases the critical temperature $T_{c0.5}$ up to 9.46 K and decreases the transition width ΔT to less than 1K at B=0 and ΔT ≤ 2.74 K at B=14 T. For the sample FeSeAg6 the activation energy for thermally activated flux flow is strongly enhanced at low fields and reaches maximal value. The significant magnetoresistive (MR) effect observed is explained by changes in the carrier concentration as a result of Ag doping. The lack of a Kohler plot in the temperature interval 30K – 90K is indicative of a multiband structure of the samples, which is in agreement with their non-Fermi liquid behavior. By analyzing the upper critical field $B_{c2}(T)$ in the framework of WHH approach, it is demonstrated that by doping of polycrystalline FeSe with Ag, a similar shape of $B_{c2}(T)$ is obtained as for FeSe single crystals for magnetic fields $H \parallel ab$ which show a flattening of $B_{c2}(T)$ at low temperatures due to Pauli limiting corresponding to a Maki parameter α = 0.92. The $B_{c2}(T)$ data at fields between 7 and 14 T exceed the orbital $B_{c2}(T)$ which is another indication for multiband superconductivity in the FeSe samples investigated. Thus, unlike in cuprates Ag addition in FeSe$_{0.94}$ not only improves intergranular connections, but significantly improves $T_c$, $B_{c2}$ and the pinning energy and enhances the MR in polycrystalline FeSe$_{0.94}$.

**Acknowledgments**

This work was conducted within the framework of the EURO Fusion Consortium, and was supported by funding from the European Union's Horizon 2020 research and innovation program, under grant agreement number 633053. The views and opinions expressed herein do not necessarily reflect those of the European Commission. This research is also partially conducted in the frame of Bulgarian-Polish Inter-academic Cooperation.

**Figure captions**

**Fig. 1** XRD patterns of FeSe$_{0.94}$ samples, doped with different Ag concentrations

**Fig. 2** Resistivity vs temperature dependences [$\rho_0(T)$] of all the investigated samples at $H=0$. Upper inset: an enlarged scale near the superconducting transition. The values of $\rho_0$ for the undoped sample are reduced by a factor of 2. Lower inset: Temperature dependences of the resistivity derivatives.

**Fig. 3** Dependence of $T_c$ vs Ag concentration at $H=0$.

**Fig. 4** Kohler plots of the FeSeAg4 sample at different temperatures.

**Fig. 5** Temperature dependences of the upper critical field $B_{c2}$ for all the investigated samples.

**Fig. 6** Temperature dependence of the upper critical field defined at $R/R_N = 0.5$ for undoped (black circles) and Ag doped FeSe samples (red circles – 4% Ag, green triangles – 6% Ag, green circles – 8% Ag). The lines mark $B_{c2}(T)$ of the WHH model for Maki parameters $\alpha = 0$ (orbital $B_{c2}$), 0.82 and 0.92.

**Fig.7** Temperature dependence of the upper critical field defined at $R/R_N = 0.5$ for Ag doped FeSe samples (red circles – 4% Ag, green triangles – 6% Ag, green circles – 8% Ag). The lines mark $B_{c2}(T)$ of the WHH model for Maki parameters $\alpha = 0$ (orbital $B_{c2}$) and $\alpha = 0.92$. For comparison, scaled $B_{c2}(T)$ data reported for a FeSe single crystal [26] are shown as blue square (see text).

**Fig. 8** ln($U/k_B$) vs ln$H$ plots for all the investigated samples.

**Figures**

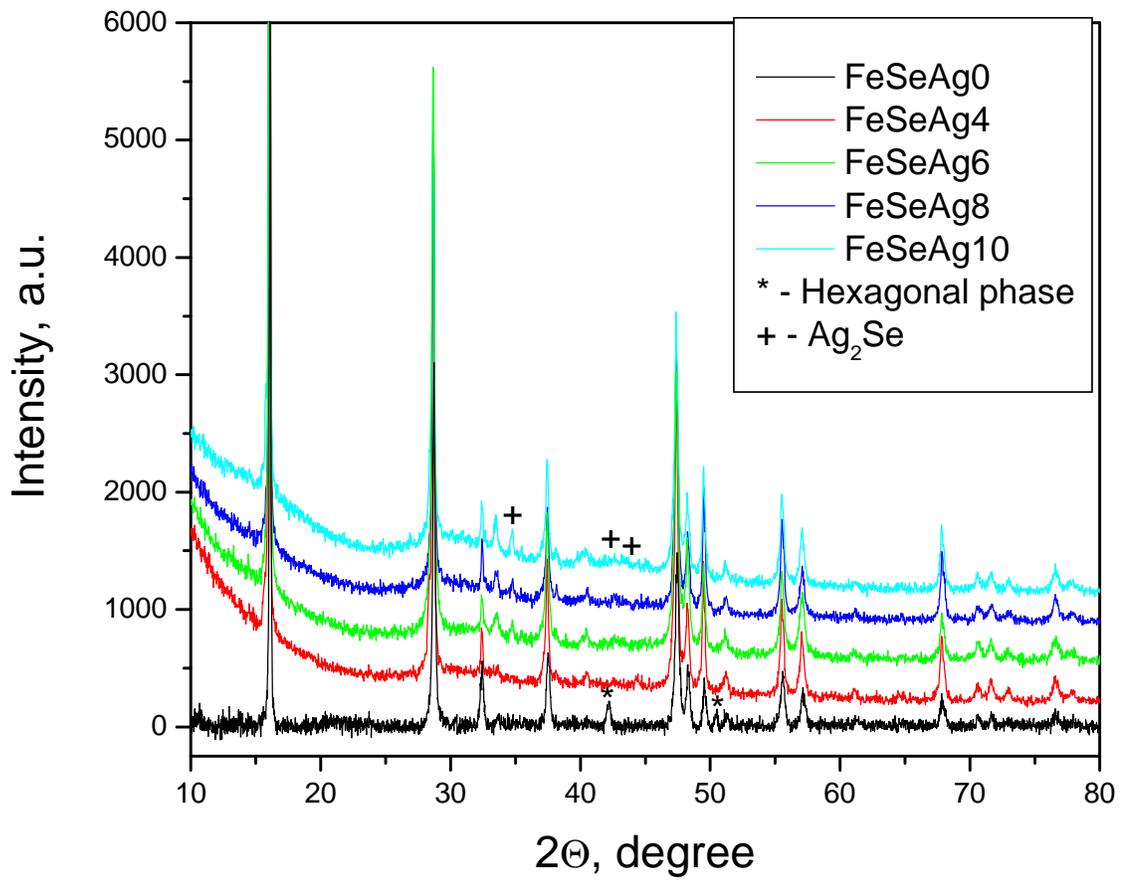

Fig.1

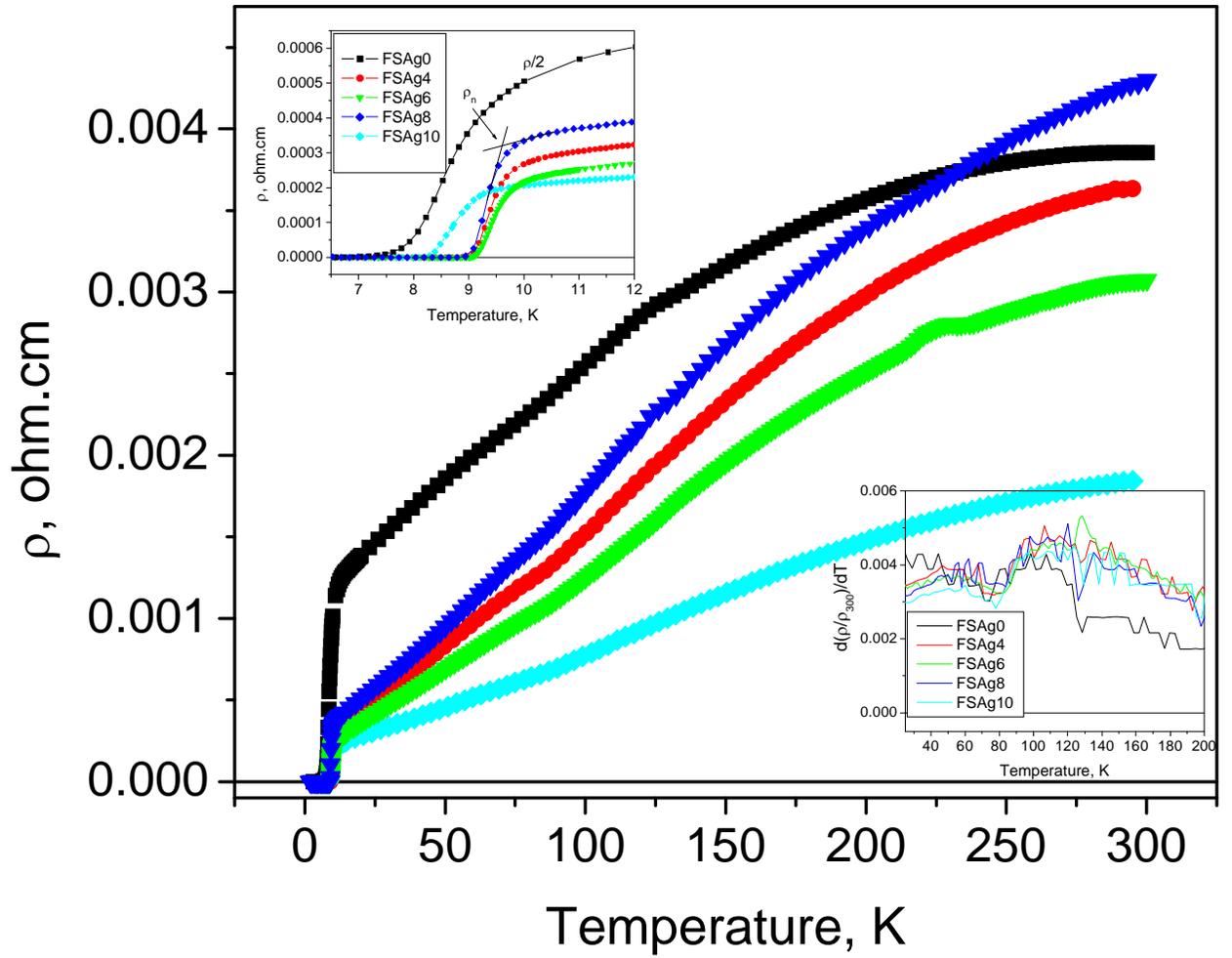

Fig.2

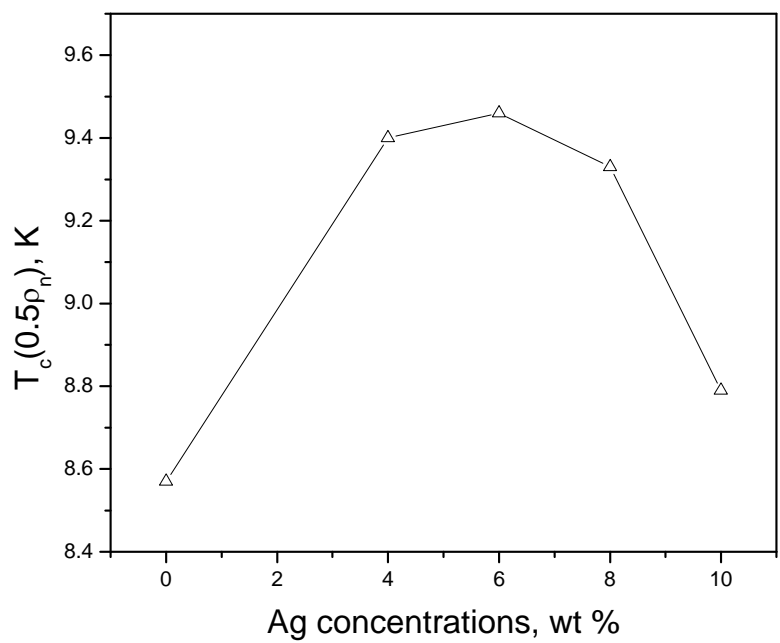

Fig.3

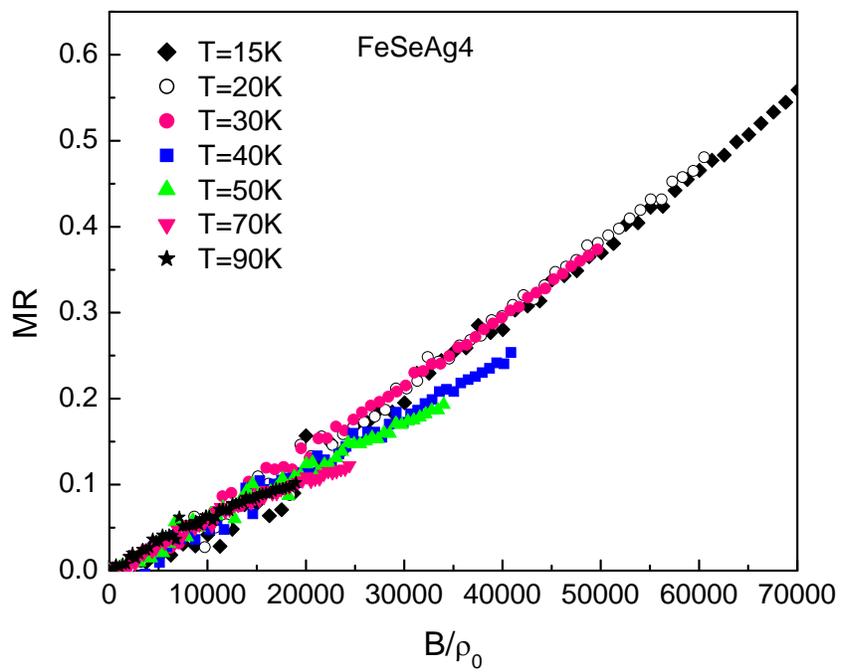

Fig.4

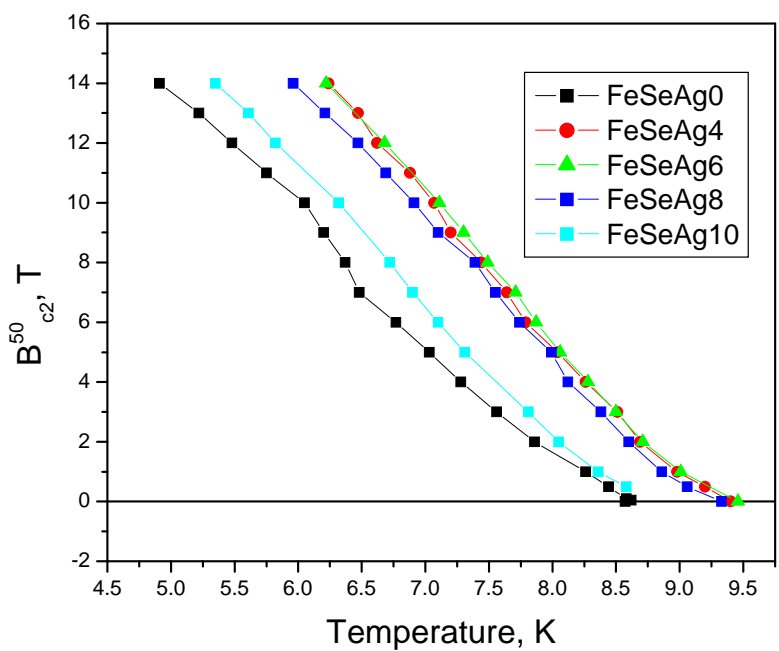

Fig.5

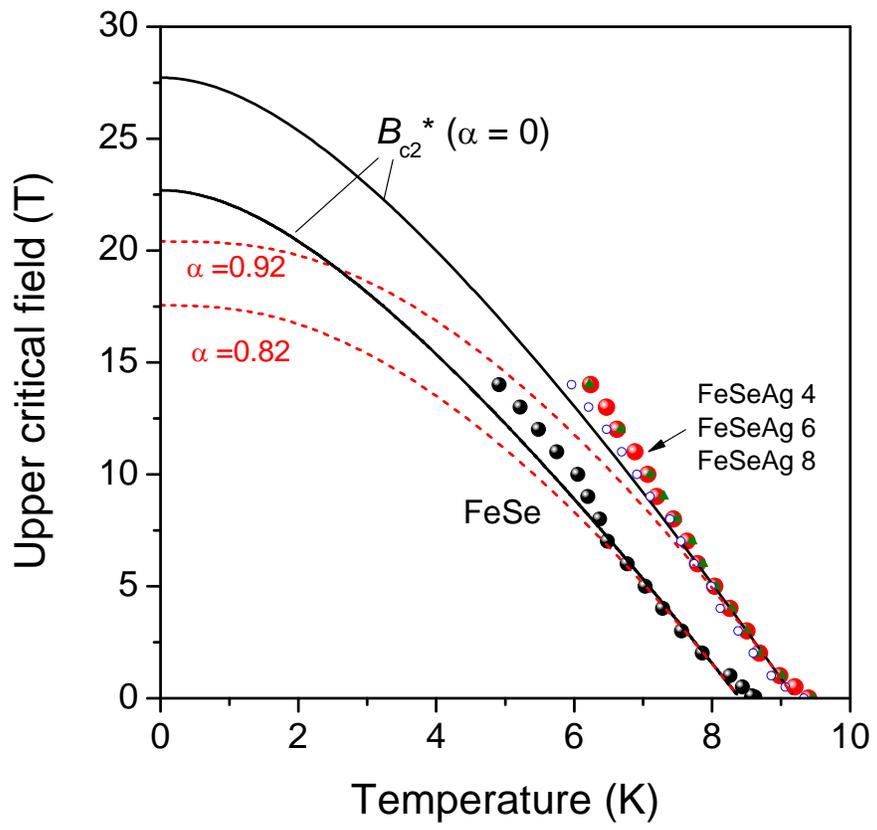

Fig.6

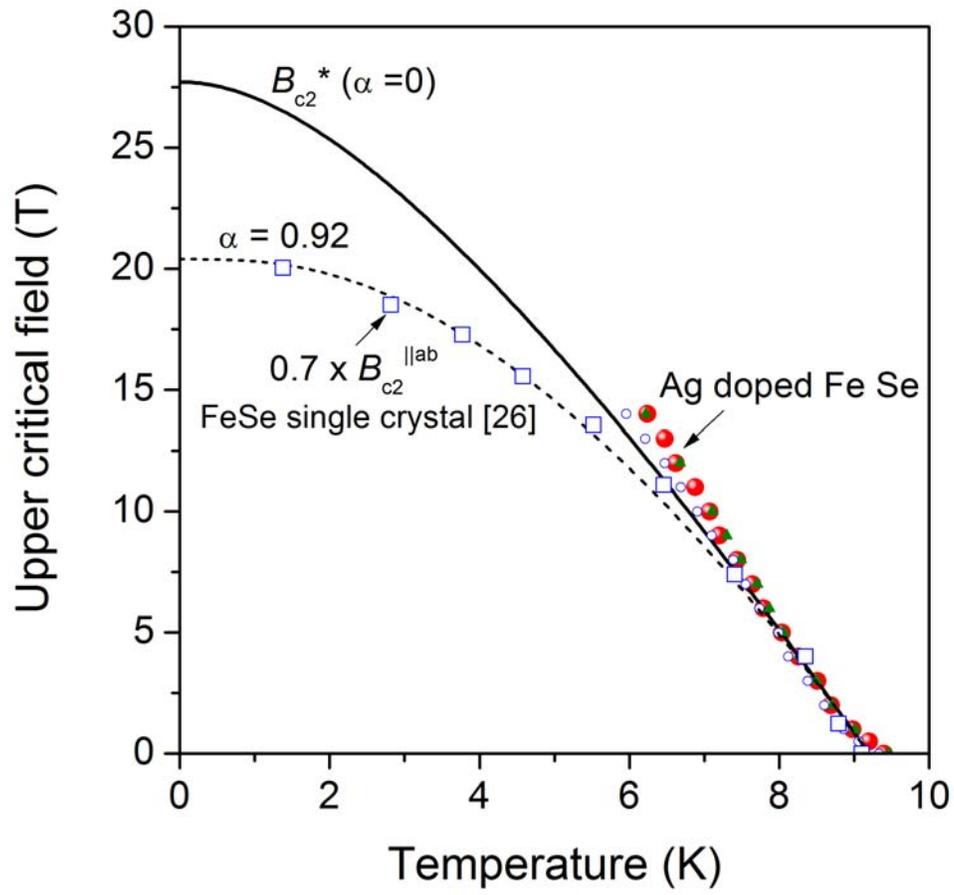

Fig.7

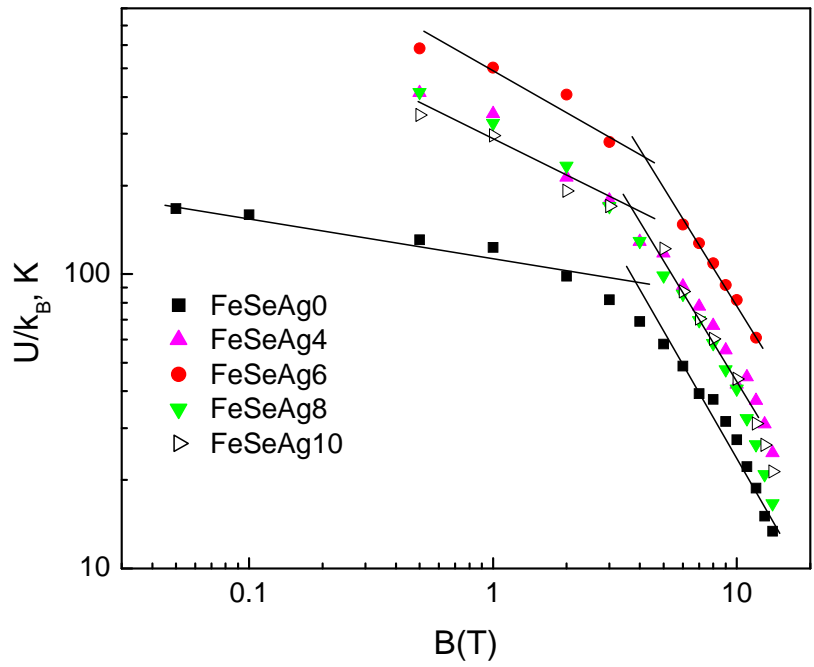

Fig. 8

Tables

Table 1. 2θ values for (00l) reflexes taken from XRD patterns for non doped sample and for two Ag doped samples (FeSe6 and FeSeAg10). Δ(2θ) is the difference between 2θ values of non doped and corresponding Ag doped sample for given (00l) reflection.

| 00l reflection | 2θ for FeSeAg0 | 2θ for FeSeAg10 | Δ(2θ) FSAg0 - FSAg10 | 2θ for FeSeAg6 | Δ(2θ) FSAg0 - FSAg6 |
|---|---|---|---|---|---|
| 001 | 16.120 | 16.028 | 0.092 | 16.052 | 0.068 |
| 002 | 32.455 | 32.383 | 0.072 | 32.414 | 0.041 |
| 003 | 49.523 | 49.473 | 0.050 | 49.492 | 0.031 |
| 004 | 67.822 | 67.806 | 0.016 | 67.838 | 0.016 |

Table 2. Superconducting parameters of the investigated samples: RRR - residual resistivity ratio; MR – magnetoresistance; ΔT - transition width, according to 10-90% criterion and $U_o/k_B$ – pinning activation energy at 1 T.

| Sample | RRR $\rho(300)/\rho(16)$ | MR (%) | ΔT(B=0), (K) | $U_o/k_B$ at 1T, (K) |
|---|---|---|---|---|
| FeSeAg0 | 2.95 | 6.4 | 2.27 | 150 |
| FeSeAg4 | 9.55 | 50.5 | 0.85 | 350.6 |
| FeSeAg6 | 9.74 | 36.4 | 0.76 | 503.2 |
| FeSeAg8 | 9.47 | 39.6 | 0.70 | 326.4 |
| FeSeAg10 | 3.25 | 34.3 | 0.93 | 295.7 |

Table 3. Superconducting critical temperatures, according to the $0.1\rho_n$ and $0.5\rho_n$ criterion; slopes $dB_{c2}/dT$ near $T_c$ of $B_{c2}^{90}$ (T) and $B_{c2}^{50}$ (T); $T_c^*$ is $T_c$ of the WHH model derived at $0.5\rho_n$; $B_{c2}^*(0)$ – orbital upper critical field at $T = 0$ according to Eq. (1); $B_p(0)$ – Pauli limiting field according to Eq. (3); Maki parameter α according to Eq. (2)

| Sample | $T_c^{10}$ (K) | $dB_{c2}^{90}/dT$ (T/K) | $T_c^{50}$ (K) | $dB_{c2}^{50}/dT$ (T/K) | $T_c^*$ (K) | $B_{c2}^*(0)$ (T) | $B_p(0)$ (T) | α |
|---|---|---|---|---|---|---|---|---|
| FeSeAg0 | 7.9 | -4.5 | 8.6 | -3.9 | 8.4 | 22.7 | 39.1 | 0.82 |
| FeSeAg4 | 9.2 | -5.43 | 9.4 | -4.35 | 9.2 | 27.7 | 42.8 | 0.92 |
| FeSeAg6 | 9.2 | -5.07 | 9.5 | -4.35 | 9.2 | 27.7 | 42.8 | 0.92 |
| FeSeAg8 | 9.1 | -5.19 | 9.3 | -4.35 | 9.2 | 27.7 | 42.8 | 0.92 |
| FeSeAg10 | 8.4 | -5.14 | 8.8 | -4.35 | | | | |